\documentstyle[11pt,aaspp4]{article}
\begin{document}
\received{ }
\accepted{ }
\lefthead{ Wang et al.}
\righthead{ }
\title{BROAD BAND X-RAY OBSERVATIONS OF THE NARROW LINE X-RAY GALAXY NGC 5506}
\author{T. WANG\altaffilmark{1,2}, T. MIHARA\altaffilmark{1}, C. OTANI\altaffilmark{1}, M. MATSUOKA\altaffilmark{1}
H. AWAKI\altaffilmark{3}}
\altaffiltext{1}{The Institute of Physical and Chemical Research ({\it RIKEN}), 2-1, Hirosawa, Wako, Saitama, 351-0198, Japan}
\altaffiltext{2}{Center for Astrophysics, University of Science and Technology of China, Anhui, 230026, China}
\altaffiltext{3}{Department of Physics, Faculty of Science, Kyoto University, Sakyo-ku, Kyoto, 606-01, Japan}
\authoremail{wang@crab.riken.go.jp}

\begin{abstract}
We present a detailed analysis of broad band X-ray data of the Seyfert 2 galaxy
NGC5506, obtained with the ASCA and ROSAT, to address the nature of Fe K$\alpha$ line 
profile and the soft X-ray excess in this AGN. Variations up to 60\% in the 
2-10 keV band are detected during a 1-day ASCA observation performed 
in January 1997, while no significant change in the 2-10 keV continuum 
shape is found. The ASCA spectrum consists of an absorbed power-law, a 
'soft excess' below 2 keV, 
and an Fe K$\alpha$ emission line at 6.4 keV. The 'soft excess' can be well 
described by either thermal emission from very low abundance material at a 
temperature kT$\simeq$0.8 keV,  or scattered/leaking flux from the primary 
power-law plus a small amount of thermal emission. The luminosity of the thermal 
emission in the former case is 1.2$\times$10$^{40}$~erg~s$^{-1}$ over the 0.5-2 keV 
band, while the excess is $\sim$ 1\% of the intrinsic hard X-ray 
continuum in the latter case. 
Analysis of ROSAT HRI data reveals that the soft X-ray emission 
is extended on kpc scales in this object, and the extended component 
may account for most of the soft X-ray excess observed by the 
ASCA. The result suggests that in this type 2 AGN, the 'soft excess' at 
least partly comes from an extended region, 
imposing serious problem for the model in which the source is partially 
covered.
We argue that the generally low abundances are a drawback for 
the thermal model, favoring a scattering dominated model.  
The scatterer is likely to be relatively cold (kT $\ll$ 1 keV)  
in this object. 
 
Fe K$\alpha$ profile is complex and can not be satisfactorily modeled by 
a single gaussian. Models of either double gaussians, or a narrow gaussian 
plus a line from a relativistic accretion disk viewed at an inclination 
of about 40$\pm$10$^\circ$ provide good fits to the data. However, the 
inclination of the  
disk can be substantially larger if there is a small amount of excessive 
Fe K edge absorption.
The intermediate inclinations for NLXGs are consistent with the ideas 
that the inner accretion disk is aligned with the outer obscuring torus.

\end{abstract}

\section{INTRODUCTION}

The discovery of weak, broad lines in the polarized light in some Seyfret 2 
galaxies lead to the general picture, the so called Seyfert unification scheme, 
that the central engine of Seyfert 1 and Seyfert 2 are essentially the same, 
and the viewing angle is the only parameter of importance in determining the 
appearance of all Seyfert galaxies (\cite{ant93}). In Seyfert 2 galaxies, 
the nuclear light is 
blocked by a thick, dusty torus from being directly observed and 
fine particles, located above the torus, scatter a small fraction of light 
into our line of sight, resulting the observed polarized broad emission 
lines (Antonucci \& Miller 1985).
In Seyfert 1 galaxies, the nuclear light is not significantly attenuated by 
this obscuring material owing to our preferential direction. However, the 
conjecture that both types of AGN have the same type of engine 
cannot be proved 
by polarized light itself, and must be explored by other means. 

X-ray observations provide a unique tool to probe the nature of   
Seyfert 2 galaxies in this respect. X-rays can penetrate material up to 
a column density of $\sim 10^{24}$~cm$^{-2}$, which allows us to directly 
view the 
very center of nuclei in at least some Seyfert 2  galaxies and, therefore, to 
make a straight test of the assumption behind the unification scheme. The 
rapid variability, Seyfert 1-type X-ray spectral chracteristics, and in 
particular, the disk-type FeK$\alpha$ line emission, should be revealed 
by X-ray observations of Seyfert 2 galaxies.
Furthermore, the amount of material along our line sight to the nucleus, 
which is an important parameter in the unification model, can be directly measured 
from the X-ray spectra. The distribution of column densities 
provides an important clue to the structure of obscuring torus. 
Finally, the fraction of scattered light in the unification model, which 
can hardly be directly measured in other band, can be determined 
technically from X-ray observations. 

To date, X-ray observations are qualitatively in agreement with the 
unification models.  Type 2 AGN are absorbed in the X-ray band by 
column densities $10^{22}-10^{24}$ cm$^{-2}$ (\cite{awa97}), sufficiently 
high to suppress the optical broad lines if they are located outside the BLR. 
Rapid variations in X-ray flux were detected in a number of Seyfert 2 
galaxies (\cite{tur97a}). Below $\sim$2 keV, emission in excess of the 
extrapolation of a  single power-law fit is usually observed, which is 
also consistent with a few percent of scattered light predicted 
by the unification model (\cite{mul93,tur97b}).
The detection of strong 
Fe K$\alpha$ emission in Seyfert 2 galaxies is also in line with 
the unification scheme (\cite{awa97}). 
 
However, it is not clear how these observations reconcile with the 
unification model in detail. Fe K$\alpha$ profiles with reasonable 
quality data have been obtained only for a few Seyfert 2 galaxies. 
Turner et al. (1998) 
examined the profiles of a sample Seyfert 2s from the ASCA archive, 
and claimed that the line profiles are similar to those of Seyfert 1's, 
 indicating reprocessing by face-on disks and contradicting the 
expectation of the unification scheme. **Contrary to this, Weaver \& 
Reynolds (1998) found intermediate inclination angle of accrestion 
disk when a narrow emission line from torus is considered.**  
The results of Turner et al. (1998) can also be due to improper continuum modeling. 
Among three NLXGs for which Turner et al. (1998) claimed with high 
confidence evidence  of face-on disks, two have unusually flat X-ray 
spectra and the third possesses a
reflection component. In view of models proposed to explain the  
flatness of the X-ray spectrum, the absorption might also be complex.
 
Although scattering by the ionized electron-scattering ``mirror'' is an 
attractive simple model for the soft X-ray 
excess, current observations showed that the situation is 
more complicated. In a number of cases, the spatally extended emission on 
kpc or even 10 kpc scales is observed (\cite{mor95,wea95,kom98}), and 
emission line 
features in the soft X-ray band are detected in a number cases (\cite{net98}) 
. These characteristics can be explained by thermal emission 
from the extended region, perhaps powered by starburst activity or an 
AGN-driven outflow (\cite{mar93}). In some 
other cases the soft X-ray emission is compact. Light from the neucleus 
that ``leaks'' through material partially covering the source is also 
a viable explanation. 

NGC 5506 is a relatively nearby bright Seyfert 2 galaxy viewed nearly 
edge-on. Broad wings on the Balmer lines and broad infrared hydrogen lines 
are claimed to have been detected (\cite{ver81}). Both the narrow and broad 
emission lines in this galaxy
are heavily reddened with $A_v=8.1$ and 60~mag respectively (\cite{vei97}). 
Minor-axis outflows from the nucleus are seen in extended optical and radio 
emission (\cite{col96a,col96b} ). Rapid X-ray variability down to time 
scales of a few 1000s was detected by EXOSAT 
(\cite{kon97}), a soft X-ray excess, fluorescent Fe K$\alpha$ line 
and evidence for either an absorption edge or a reflection bump was 
found in the Ginga data (\cite{bon93}). Furthermore, extended soft X-ray 
emission aligned with the minor axis outflows is observed (\cite{col98}). 
These authors interpreted the extent as thermal X-ray emission from the 
outflow plasma.  These properties, together with 
its brightness in the X-ray band, make NGC 5506 an ideal target for tackling all these 
questions. In this paper, we present a detailed analysis of the X-ray spectrum 
and image from ASCA and  ROSAT HRI observations of this object.

\section {OBSERVATIONS AND THE DATA REDUCTION}

NGC 5506 was observed by the ASCA on 1997 January 30. The observation was carried 
out in the 1CCD faint 
acquisition mode for the two SIS detectors and in the PH mode for the two GIS 
detectors. 
The faint mode data were converted into bright2 mode data, which corrects 
then for the
dark frame error and echo effects. After the data were cleaned using 
standard screening criteria, the remaining usable time was 43.3 ks for the GIS 
and 39.2 ks for the SIS. 
Hot and flickering pixels were removed from the SIS data and rise-time 
rejection was applied to exclude particle events for the GIS data.
SIS photon event grades 0, 2, 3, 4 were selected for the analysis. The source counts were 
extracted from circular regions of radius~$\sim$3.5\arcmin~ and $\sim$7\arcmin~ for the 
SIS and GIS, respectively. The background counts for the GIS were 
extracted from a region near the source. For the SIS, the background 
spectra were estimated from a source free region at the same chip of the CCD. 

After correcting for the background, which accounts for 2\% to 4\% of the 
total counts, the source count rates are 0.941$\pm$0.005, 0.727$\pm$0.004, 
0.800$\pm$0.005, 0.880$\pm$0.005 cts/sec in the 0.5-10 keV range for SIS0 and SIS1, and  
in the 0.7-10keV range for GIS2 and GIS3, respectively. The data preparation 
and spectral analysis were performed using version 1.4 
of the XSELECT package and version 10.01 of XSPEC.

In addition to the ASCA observation, NASA/GSFC archival ROSAT HRI data 
were used for the image analysis. The HRI observation was 
performed in 1994, 3 years ahead of the ASCA observation. 

\section{RESULTS}

\subsection{Time variability}

Figure 1 shows the 2.0-10.0 keV source+background light curves binned at intervals of
256 s for the GIS and SIS. The background is negligible and, 
therefore has not been subtracted. The lightcurve consists 
of two major events. At about $\sim 2\times 10^4$ s after the start of 
the observation, the source brightens by $\simeq$30\% higher and then 
decreases to its begining level in 
$10^4$ s. The source increases by about 20\% in  
$2\times 10^4$ s, and then decreases during again.  
The ratio of the maximum to minimum count rate is 1.7, or a maximum variabilty 
amplitude of 70\%. Variations up to 10\% are visible down to the time scale 
as short as 1000 sec. The rms fraction variation is estimated to be 
0.13$\pm$0.01 and 0.11$\pm$0.01 for the 256 s binned SIS and GIS light curves,
respectively.  

Soft (0.5-1.0 keV) and hard (2.0-10.0 keV) band light curves for SIS data 
have been examined in order to see if the fluxes in both bands vary 
differently. The curves were binned at intervals of 5760s in order to 
increase S/N ratio of each bin in the soft band. A background light-curve, 
which has been extracted from a source free region and binned at the same 
intervals, is subtracted from the source light curves.
 The count rates in the soft band are consistent with no variations 
($\chi^2_\nu=0.7$, for 16 degrees of freedom). However, 
a model in which the count rates at the soft and the 
hard band are proportional also gives an acceptable fit ($\chi^2_\nu=1.08$ 
for 16 degrees of freedom), although the $\chi^2$ is worse by 6. Therefore 
this analysis is inconclusive due to low statistics.   

Spectra from the high and low states (marked in Fig. 1) have been accumulated 
from the data. 
Fitting a power-law plus a broad gaussian Fe K$\alpha$ model (see below) to 
 both spectra yields similar spectral indices ($\Gamma_l$=1.78$_{-0.04}^{+0.06}$ 
and $\Gamma_h$=1.80$_{-0.05}^{+0.06}$ for the low and high state, 
respectively), line energy ( E$_l$=6.44$\pm$0.07 and 
E$_h$=6.34$\pm$0.10 keV), equilvalent width (EW$_l$=190$_{-40}^{+60}$ eV and 
EW$_h$=220$_{-70}^{+80}$ eV), and line flux ($f_l$ = 1.57$_{-0.35}^{+0.53}$
$\times$10$^{-4}$ and $f_h$ = 1.82$_{-0.40}^{+0.66}$$\times$10$^{-4}$ photon~cm$^{-2}$~s$^{-1}$).

\subsection{Spectral Properties}

The 2-SIS spectra and the 2-GIS spectra 
 are combined together to improve statistics, while fits to GIS 
and SIS spectra are preformed separately. The spectra have been regrouped to have 
a minimum of 25 counts per energy bin. In the following, all errors are quoted 
at $\Delta\chi^2=2.71$, or 90\% confidence level for single free parameter.
Energies are referred to the observer's frame 
unless specified. A column density of $4.2\times 10^{20}$~cm$^{-2}$ is adopted 
for the Galactic absorption (\cite{elv89}).  

The broad band (0.5-10.0 keV) spectrum is complex, showing a soft X-ray 
excess and an Fe K$\alpha$ emission line with respect to a single absorbed 
power-law  fit. In the subsequent sections, we will present detailed 
modeling for the Fe K$\alpha$ line and the soft excess. 

\subsubsection{The Fe K$\alpha$ Line Profile}

Before describing the detailed modeling, we show the Fe K$\alpha$ line 
profile in Figure 2. A strong narrow core and complicated broad wings 
are clearly indicated in the plot. 
Since Figure 2 shows the ratio of data to the model and the 
continuum is weaker at higher energy, the actual profile should peak 
at somewhat lower energy.

To examine tha Fe K region, only the data above 2.0 keV were fitted. This  
avoids complications 
of modeling the soft excess. The continuum is modeled as a single 
power-law absorbed by a fully covered material with the cosmic abundance. As a 
consistent check, including the soft excess component does not significantly 
affect the Fe K$\alpha$ fit. The results are presented in Table 1 and 
Table 2. 
 
The line profile is sensitive to the exact placement of continuum, and so 
a careful modeling of continuum is required. Although a reflection  
component was suggested by the Ginga data (\cite{bon93}),  
adding it to the models described below does not improve the fit, and 
the covering factor always converges to a value close to zero. 
Therefore we only describe the power-law continuum model. And further, 
SIS and GIS fits yield consistent results, therefore we fit the SIS and GIS 
data simultaneously in order to better constrain the parameters, allowing 
the normalizations of the power-law component to vary independently. The 
Results are listed in Tables 1 and 2. 

Initially, the line is modeled with a narrow gaussian ($\sigma$ is fixed 
at 0.02 keV). The fit yields a line energy E=6.30$\pm$0.06 keV and 
equivalent width of the line 134$\pm$37 eV. Although the fit is 
statistically acceptable ($\chi^2_\nu/\nu$=1.041/1105, see Table 1), 
the systematic deviations of the data from the model suggest that 
the fit is poor. Excesses to the blue and red sides of the line are 
visible in the residuals (Figure 3). 
The fit is improved at $>99\%$ confidence level, with a $\Delta\chi^2=
18$,  when the line width is 
allowed to vary. The best fitting line width is
0.20$_{-0.11}^{+0.21}$ keV. The broad line fit yields  
significantly more line flux, with an EW of 203$_{-74}^{+94}$ eV, because 
the broad line fit picks up the flux in the line wings.
 
Structure is visible in the residuals and so we try a complicated model. 
A model with a broad plus a narrow gaussian is fitted to the data, and the 
residuals is shown in Fig. 3b. 
The fit is improved at a confidence of $>$95\% level, with $\Delta\chi^2\simeq$9 
for 2 more degrees of freedom, relative to the broad gaussian fit. The broad line 
is found at energy of 6.3$_{-0.3}^{+0.2}$ keV, slightly redshifted relative 
to the narrow one, with a width of $\sigma$=0.6$_{-0.3}^{+0.5}$ keV (Table 2).
 The line fluxes in the narrow and broad components are similar. 
In comparison with the  
line profile of the Seyfert 1 galaxy MCG 6-30-15 (\cite{iwa96}), the 
broad line 
in NGC5506 is less redshifted (E$_b$=5.9$\pm$0.1keV for MCG 6-30-15), 
has a similar width ($\sigma=0.6$ keV) and is weaker (EW$_b \simeq 200-400$ eV).   

Next, we model the line assuming emission from an 
accretion disk around a Schwarzschild (\cite{fab89}, hereafter Fabian; 
Fig. 3c ) or extreme Kerr black hole (\cite{lao91}, Laor; Fig. 3d ). 
The parameters are the 
inner and outer accretion disk radii, the index of disk emissivity law 
q ($F\propto r^{-q}$), line intensity, and the inclination of the system. 
Since the line has a strong 
narrow core, the outer radius of the disk must be large. 
It should be mentioned that the Loar91 model in the current XSPEC does not 
calculate the disk emission beyond 400 r$_g$. 
In all fits presented below, the outer radius is allowed to vary freely with 
upper limit of 400 r$_g$ for Laor model and 10$^6$ r$_g$ for Fabian model,  
while the inner radius is fixed at 1.235 and 6 r$_g$ for Loar91 and Fabian 
models respectively. The best fits yield to a q=2.2$\pm$0.1 and 
1.8$^{+0.4}_{-0.8}$, and inclinations 16$_{-16}^{+10}$$^\circ$ and 
41$\pm$10$^\circ$ for the Laor and Fabian models, respectively. 
The different results can be simply due to the fact that the outer radius 
of Laor model does not go beyond 400 r$_g$. We have noticed that 
for given outer and inner radii, the the inclination is mainly 
determined by the blue peak energy, which moves to a low value when the 
outer radius increases except for extremely face-on disk and at very small 
radius (Fig. 1 of Laor 1991). So taking the emission line from outside 
400 $r_g$ falsely as emission wthin it will result in an apparently low 
inclination. This is exactly the case we did here with Laor model. The best 
fiting outer radius is $10^{5}$ $r_g$ 
for Fabian model and reached the upper boundary for Laor model. 

Keeping in  mind that the absorbing material also emits 
Fe K$\alpha$, we add a narrow gaussian to the disk models 
(Fig. 3e and Fig. 3f).  
The line energy and outer radius of the disk are fixed at 6.4 keV and 
400 r$_g$. The fit is improved at the 90\% confidence level for 
the Laor model ($\Delta\chi^2=3.4$ for 1 more constraint), while 
no significant improvement occurs for the Fabian case. This time, however, the line parameters 
for the two models are similar, with an inclination for the disk 
of 40$\pm$10$^\circ$ for both models. The EW of narrow line is 30-80 eV for 
the Fabian model and 50-110eV for the Laor model. The disk line has an 
equivalent width 
of approximately 200 eV.

Inspecting the residuals, we find an excesses between 7.7-8.3 keV (Fig. 3e,f).
This might be part of the blue wing, if there is an absorption edge at around 
7 keV. An absorption edge might be expected if the Fe is 
over-abundant in the absorbing material or the absorption is complex.  
To explore this possibility, we added an Fe K edge to the model,
The fit is moderately improved for the Loar91 model, with a $\Delta\chi^2$=4.4 
for 2 more free parameters. For this case, the best fit converges to 
i=69$_{-9}^{+20}$$^\circ$ and $\tau$=0.10$\pm$0.07.  
Fot the Fabian model, the improvement 
is not significant.  

\subsubsection {The Soft X-ray Excess} 

Next we try to model the soft X-ray band. As shown in 
the last section, the detailed spectral slope and the absorbing column 
density are not sensitive to the modeling of Fe K$\alpha$. Here 
the Fe K$\alpha$ line is modeled with double gaussian (see above) and 
the line 
energies and the widths are fixed at their best fitting values (Table 1). 
The primary continuum is described as an absorbed single 
power-law, with neither reflection nor absorption edge included.

The soft excess is modeled as either partially absorbed optically thin plasma 
emission ( Raymond-Smith model; Fig. 4a ) or a power-law (Fig. 4b). In the 
latter case, the spectral index is fixed at 
the value for the primary one; this corresponds to the case of 
partially covered absorption or scattering of the primary continuum (refered to 
as scattering model below). 
The results are summarized in Table 3. An excess is still visible between 
0.8-1.0 keV (Fig. 4 (a) and (b)) for both models, although the thermal 
model produces 
a better fit than the scattering model. The total luminosity 
in thermal component in the 0.5-2 keV band is $1.2\times 10^{40}$ 
erg~s$^{-1}$ (assuming H$_0$=50), 
the abundances are extremely low with Z = 0.03$_{-0.01}^{+0.04}$Z$_\odot$,
and the best fit plasma temperature is $\simeq$
0.8 keV.  For the scattering model, the normalization is $\simeq$1.3\% of the 
primary unabsorbed component, which corresponds to a partial covering factor 
$\simeq$98.7\% in the case of partially covered absorption. 

For the scattering model, the excess between 0.8-1.0 keV  
could be a thermal emission component. If the scatter is the 
hot plasma, it must physically produce thermal emission as well. 
When a Raymond-Smith component (abundance fixed at solar value) is 
added to the scattering model,
the fit is dramatically improved with a $\Delta\chi^2$=17 for 2 more 
degrees of freedom ( Fig 4 (c)), but the amount of thermal emission is 
uncertain (Table 3). 
This fit has the same number of parameters as the free abundance 
Raymond-Smith model, and produces a slightly better fit with $\Delta\chi^2 
= 6.5$. 

\subsection {Spatial Extent at HRI Image}

The ROSAT HRI image has been analyzed by Colbert et al. (1998), who  
estimate the X-ray emission to be extended on the kpc scale. We re-analyze 
the HRI data to determine if the soft X-ray excess seen with ASCA
can be explained by the extended emission. The best 
fitting soft excess model (RS+PL in Table 4) for the soft X-rays 
convolved with the ROSAT HRI response predicts 
a count rate 0.024 cts~s$^{-1}$; 0.008 cts~s$^{-1}$ for the RS component 
and 0.016 cts~s$^{-1}$ for the scattered component. 

The HRI image is extracted from the events file obtained from the ROSAT 
archive at GSFC. Using XIMAGE, we estimate source count rate of 
0.049$\pm$0.002 cts~s$^{-1}$, which is 
in good agreement with that obtained  by Colbert et al. (1998).
Background accounts for $\sim$ 10\% in a 1\arcmin~ circle. The radial 
profile of counts per unit area and the in-flight PSF, binned at intervals 
of 0\arcmin.05, are plotted in Figure 5. 
Excess emission beyond 0\arcmin.1 
is obvious with weak emission extending up to 0\arcmin.4. 
The count rate from the extended emission can be estimated by subtracting 
the normalized PSF from the radial profile of NGC5506 and integrating 
between 0\arcmin.1 to 0\arcmin.4. This yields a net count rate from the 
extended 
component (beyond 0\arcmin.1), $\sim$ 0.015$\pm$0.002 cts/s, or 25 to 35\% of 
the total HRI counts. This count rate is slightly lower, but approximately 
consistent 
with that derived from the soft X-ray excess model from the ASCA band.
However, we are cautious about this result, since an uncertainty in 
the aspect solution occasionally produces an apparent source extent 
in the HRI observed sources (\cite{dav92,bri94}) which was found to be 
of the order 1\arcsec~ in FWHM (\cite{mor94}). 

Colbert et al.'s conservative estimation of 
0.0040$\pm$0.0008 cts~s$^{-1}$ for the count rate between 10\arcsec~ and 
25\arcsec, is about 8\% of total HRI count rate, and only accounts 
1/6 of the ASCA soft excess derived from the ASCA spectral fitting.
And this number is also a factor of 2 lower than the contribution from 
RS component in the RS+PL model (Table 4).

\section {DISCUSSION}

We have shown that the soft-X-ray excess in NGC 5506 is spatially extended 
on kpc scales and can be fitted by thermal plasma emission with very low 
abundances having temperature kT$\simeq$0.8 keV, or by a scattered primary 
power-law component plus a small amount of thermal emission with cosmic 
abundances. 

Although the luminosity, $\simeq$10$^{40}$ erg s$^{-1}$, and the temperature 
derived from the optically thin thermal plasma model is well within the range found 
for normal galaxies (\cite{mat96}), the extremely low 
abundances however are a drawback. The thermal emission from late type galaxies
usually has abundances of only a few times lower than the solar value, but 
we find  a factor of 50 times lower. This is unexpected because the optical 
narrow line spectrum 
of NGC5506 is similar to the other Seyfert galaxies; and the ratio of metal 
lines to the Balmer lines are of typical value (\cite{mal86}).
Very low abundances have been found in some other type 2 AGN when a single 
 component thermal 
model was fitted: 0.07$_{-0.05}^{+0.11}$ Z$_\odot$ for NGC2110 (\cite{hay96}),
 0.05 Z$_\odot$ for NGC4388 (\cite{iwa97}). 
The low abundances imply that the emission lines are weak in comparision 
with the continuum emission. This can be due to the extra emission in the 
continuum, e.g., from plasma with temperature much higher (e.g. $>$2 keV) 
than the one produce the line features in the 0.7-1.0 band. High ionized  
plasma is insufficient line emitter.  In order to account for so low 
abundances, the high temperature gas should be a factor 10 more than the 
low temperature, but it still emit only small amount of total flux in 
2-10 keV band. Another possible solution is that the gas is much higher 
ionized than indicated by its excitation level (temperature), this is a 
characteristic of photionized gas. Simulations is needed in order to see 
if these can indeed explain the low abundance.  
 
If the soft excess is scattered continuum flux, the scattered light accounts 
for 
$\sim$1\% of unabsorbed primary power-law. The scattered fraction  
is typical for type 2 AGN observed by ASCA (\cite{tur98}). 
However, the actual fraction is hard to estimate since the scattering 
region is likely to be spatially extended on kpc scales and the AGN 
is variable.

If the emission between 6\arcsec -10\arcsec~ is real, then the size 
of the scattering region is of order of  $R\sim$1 kpc.
An order of magnitude estimate would suggest that the scatterer is 
relatively cold (kT$<<$1keV), and that the extra thermal emission is only 
from a small amount of gas, perhaps hot phase gas. For an optical depth 
of $\tau_{sc}\sim 0.01$, we derive a Raymond-Smith normalization of, 
\[
A_{RS} = (10^{-14}/4\pi D^2)\int{n_en_h}dV
  \simeq 10^{-14}/3D^2 (\frac{\tau_{es}}{\sigma_T})^2 R^{-1}
  \simeq 0.6
\]
for a homogeneous medium, and it would be larger for other cases.
However, our spectral fitting results suggests a value of 
only $10^{-4}$ for the RS component, 4 orders of magnitude lower than 
expected. A simple simulation 
shows that the temperature of this plasma should be lower than 0.1 keV in 
order not to produce too much excess emission at low energies. Since 
$A_{RS}$ is inversely proportional to $R$, this conclusion is even valid for 
R as small as 10 pc.

  
Extranuclear scattered light on the kpc scale had not been 
detected in the X-ray, however, it was observed in the optical band in 
recent years. 
Shields \& Filippenko (1996) reported the first case of extended 
(10\arcsec) broad line emission in NGC 4388, which was interpretted 
as scattered light coming from an obscured Seyfert 1 nucleus. Ogle et al. 
(1997) found broad polarized emission in Cygnus A, and they resolved the 
polarized light into three extended regions, concluding that the broad 
emission lines arise from the scattered light of a quasar of modest 
luminosity hidden in the nucleus of the galaxy. More recently, Mediavilla 
et al. (1998) find a broad emission line region of 8 kpc 
scale in the Seyfert 1 galaxy Mrk509. Thus it is no surprise that we
 detect scattered 
light in the X-ray band on kpc scale. Future spatially resolved 
spectroscopy 
in the X-ray, optical bands should verify kpc scale emission.

The fluorescent iron line provides a diagnostic tool with which to probe 
the region immediately around the putative black hole. The line profiles 
from Seyfert 1 galaxies show a broad red wing extending to the energy 
4-5 keV (\cite{nan97}), which can be naturally interpreted as emission 
from a face-on relativistic accretion disk around a black hole (\cite{tan95}) 
, although an alternative model may also explain the characteristics of the 
Fe K line   
(\cite{mis98}). In Seyfert 2s, the situation is  somewhat 
complicated. 
For many of them, the obscuring material is so optically thick that only 
the scattered light is seen, such as for NGC1068, and their spectrum is 
dominated by a compton-thick reflection component with a very large EW of 
Fe K$\alpha$ emission. A line component similar to  Seyfert 
1s has also been identified in some cases (\cite{tur98}). For a particular 
subclass of type 2 galaxies, called NLXG, the absorbing column is less 
than 10$^{23}$~cm$^{-2}$ and is transparent to the X-rays at energies 
$>$3keV. These sources possess Fe K$\alpha$ profiles  
similar to those seen in Seyfert 1s.

For a sample of six NLXGs, Turner et al. (1998) compared the observed line 
profiles to disk-lin model preditions and 
claimed that the Fe K$\alpha$ profiles are indicative of 
reprocessing from face-on disks. The inclinations were well constrainted 
to less that 30$^o$ for 3 objects (NGC2110, NGC526A, MCG 5-23-16) among 6. 
However, Weaver et al. (1997) found an intermediate inclination 
solution for MCG 5-23-16 when a reflection component and narrow Fe K line 
was included in the model. The existence of a reflection has subsequently 
been confirmed by data from  
RXTE (\cite{wea98}). 
The remaining two objects, NGC2110 and NGC526A, show flat 
spectra in the ASCA band. It has been proposed that the absorption 
in these AGN is complicated (e.g, \cite{hay96,cap98}). 
When a complex absorption model, which consists of a partially covered 
high column and a fully covered low column densities material,   
 is applied to the data, we find the 
disk inclinations: 65$_{-22}^{+7}$$^\circ$ and 52$\pm$15$^\circ$ for 
NGC2110 and NGC526A, respectively.    
The different inclinations is due to the fact that either reflection 
component or complex absorptions contain an edge around 7.1 keV, which make 
the apparent blue wing of the Fe K$\alpha$ weaker, resulting a low 
inclination disk. When the edge is corrected, the disk inclination 
become higher. Using a 2-component (Gaussian plus disk line) model for Fe K$\alpha$~, Weaver and Reynolds (1998) also find large inclinations. An intermediate inclination (40-50$^\circ$) disk is also 
found for the type 2 AGN IRAS 18325-5926.
For NGC5506, the best fit inclination is 40$\pm$10$^\circ$, which is 
 intermediate. Based on carefully calculated numbers 
of Seyfert 1, Seyfert 2 and intermediate Seyfert galaxies, it was proposed 
that  partially obscured type 2 AGN, such as NLXG, are 
viewed at an angle between 35-55$^\circ$ with respective to the axis of 
torus (\cite{ost88,ost93,mai95}). The intermediate inclinations derived for 
these objects are consistent 
with ideas that  the inner disk and outer torus are aligned, contrary to Turner et al.. 

The narrow Fe K line in our two component fit has an EW of $\sim$ 70eV, which is 
comparable to the same component in 4 NLXGs (\cite{wea98}). 
The line EW is still somewhat larger than predicted 30 eV from 
absorbing material with the observed column density uniformly covering 
the source at cosmic abudances.   
However, if we view through the low column density part of the 
obscuring torus, and the majority of the torus is much thicker, as 
proposed for partially obscured type 2 AGN (\cite{vei97}), the 
torus may well explain the narrow line component. Alternatively, a large 
EW may also result from an over-abundance in Fe. 	

\section {CONCLUSION}

We have analyzed the broad band X-ray data of NGC5506 collected from the ASCA 
and ROSAT mission. The main results can be summarized as follows:\\
\begin{enumerate}
\item Up to 60\% variations in the 2-10 keV count rate are observed during 
a 1-day 
ASCA observation. The ASCA spectrum consists of an absorbed power-law, 
a 'soft excess' below 2 keV, and an Fe K$\alpha$ emission line at 6.4 keV. 

\item The soft excess can be well
described by either thermal emission from extremely low abundance material at a
temperature kT$\simeq$0.8 keV  or a scattered/leakage of a primary
power-law plus a small amount thermal emission. For pure thermal, we 
find $L_x\simeq 1.2\times10^{40}$~erg~s$^{-1}$ in the 0.5-2 keV
band, while if we assume the soft X-rays are scattered, the excess is 
$\sim$ 1\% of the directly viewed hard X-ray
continuum.

\item ROSAT HRI data reveal that the soft X-ray emission is extended on a 
kpc scale in this object. The extended component may account for most of 
the soft X-ray excess observed by the ASCA.   

\item The Fe K$\alpha$ profile is complex, and can not be satisfactorily modeled by
a single gaussian. Models of either double gaussians, or a narrow gaussian
plus a line from a relativistic accretion disk viewed at an inclination angles 
about 40$\pm$10$^\circ$ provide good fits to the data. The inclination could be 
larger if there is small amount of excess FeK edge.
\end{enumerate}

We conclude that the soft X-ray excess is most likely dominated by  
 nuclear X-rays scattered from relatively cold material, plus a small 
amount of thermal emission from extended thermal matter.  The later can be 
identified with the large scale outflow gas by (Colbert et al. 1998). Future 
spatially resolved spectroscopy in X-rays and optical can verify or dimiss   
this explanation. The intermediate disk 
inclinations found for this and other NLXGs are consistent with the idea 
that the inner accretion disk and outer obscurer are aligned.

\acknowledgements

It is a pleasure to thank Taro Kotani, Hitoshi Negoro and Hidetoshi Kubo for 
useful discussions regarding the ASCA data processing. We thank refree 
Kimberly Weaver for many useful comments, which greately improve the 
presentation of thsi paper.  T.W. acknowledges 
support at RIKEN by a Science and Technology Agency fellowship. We thank 
ASCA teams for succefully carrying out this observation and data reduction. 
This research has made use of the NASA/IPAC Extragalactic Database(NED), which is operated 
by the Jet Propulsion Laboratory, California Institute of Technology, under 
contract with the National Aeronautics and Space Administraction.

\clearpage
\figcaption[fig1.eps]{ASCA SIS and GIS light curves in the 2-10 keV 
band. The count rate have been binned to 256 seconds, and 2 SIS (GIS) 
data are combined together. The background count rate, which accounts 
for 2-5\% of the total, has not been subtracted. The time is from 
MJD 128848448.9678306 sec.}

\figcaption[fig2.eps]{Ratio of the data to an absorbed power-law model fitted 
from 2 to 4.5 keV and from 7.5 to 10 keV bands to illustrate the Fe K line 
profile for the SIS (a) and the GIS (b). The data from 2 detectors were 
combined in each case. }

\figcaption[fig3.eps]{Plot of data to model ratios for the simultaneous 
fits to the SIS and GIS data over 2-10 keV.  The corresponding model for 
Fe K$\alpha$ is labeled in each panel (see also Table 1 and 2). }

\figcaption[fig4.eps]{Simultaneous fit over broad band ASCA spectrum of 
  the SIS and GIS. The Fe K$\alpha$ has been modeled with double gaussians 
and the soft X-ray excesses are fitted with Raymond-Smith emission in panel 
(b), scattered power-law in panel (c) and a scattered power law plus 
Raymond-Smith emission with cosmic abundances in panel (d). We have 
also show the fit without modeling soft excess in panel (a).}

\figcaption[fig5.eps]{The spatial profile of X-ray image around NGC5506 
from the ROSAT HRI data. The count rate, together with its error bar, at 
each radius is shown. The on-orbit HRI PSF is overploted as a solid 
line. At the distance of NGC 5506, the scale 0\'.1 corresponds to a 
physical size of 0.7 kpc.}

\newpage 

\begin{deluxetable}{cllllllllll}
\tablewidth{0pt}
\tablecaption{Gaussian fits for the Iron Line (2-10 keV)}
\tablehead{
\colhead{fit}  & \colhead{N$_H$} & \colhead{$\Gamma$} 
& \colhead{A$^{a}$}  & \colhead{E$_n$} & \colhead{EW$_n$} & \colhead{E$_b$} & \colhead{$\sigma_b$}&  \colhead{EW$_b$} 
& \colhead{$\chi^2_\nu/\nu$}  \\
\colhead{} & \colhead{} & \colhead{10$^{22}$cm$^{-2}$} & \colhead{} &
\colhead{10$^{-2}$} & \colhead{keV} & \colhead{eV} & \colhead{keV} & \colhead{keV} 
& \colhead{eV} & \colhead{} 
}
\startdata
NL  &  3.01$\pm$0.20 & 1.75$\pm$0.05 & 2.4$\pm$0.2 & 6.30$\pm$0.06 & 134$_{-37}^{+37}$ & & & & 1.041/1105 \\
BL  &  3.07$\pm$0.25 & 1.78$\pm$0.09 & 2.5$\pm$0.2 & & & 6.39$_{-0.08}^{+0.09}$ & 0.20$_{-0.11}^{+0.21}$ & 203$_{-74}^{+94}$ & 1.025/1104 \\
NL+BL  & 3.21$\pm$0.14 & 1.82$\pm$0.04 & 2.7$\pm$0.2 & 6.41$_{-0.05}^{+0.03}$ & 101$_{-37}^{+19}$ & 6.25$_{-0.25}^{+0.21}$ & 0.63$_{-0.28}^{+0.45}$ & 126$_{-50}^{+88}$ & 1.018/1102

\enddata
\tablenotetext{a}{SIS normalization at 1 keV in unit of photon~cm$^{-2}$s$^{-1}$keV$^{-1}$. }
\tablenotetext{}{fits: NL-- narrow gaussian, BL -- broad gaussian, NL+BL -- narrow plus broad gaussians.}
\end{deluxetable}

\begin{deluxetable}{cllllllllll}
\tablewidth{0pt}
\tablecaption{ Disk Line model Fitting to the Fe K$\alpha$ line }
\tablehead{
\colhead{fit} & \colhead{N$_H$} & \colhead{$\Gamma$} & \colhead{A$^a$} & \colhead{$E_n$} & \colhead{EW$_{n}$}  
 & \colhead{$E_d$} & \colhead{q} & \colhead{\it i} 
& \colhead{EW$_d$} & \colhead{$\chi^2_\nu/\nu$}  \\ 
\colhead{ } & \colhead{$10^{22}$cm$^{-2}$} & \colhead{} & \colhead{10$^{-2}$} & 
\colhead{keV} & \colhead{eV} & 
 \colhead{keV} & \colhead{}&\colhead{} & \colhead{eV} & \colhead{}
}
\startdata
Fabian  & 3.18$\pm$0.13 & 1.81$\pm$0.04 & 2.6$\pm$0.2 &  \nodata & \nodata & 6.38$_{-0.04}^{+0.04}$ & 2.2$_{-0.1}^{+0.1}$ & 41$_{-10}^{+9}$ & 225$_{-39}^{+34}$ & 1.017/1102 \\ \nl
Laor  & 3.16$\pm$0.13 & 1.80$\pm$0.04 & 2.6$\pm$0.2 &  \nodata & \nodata & 6.5$\pm$0.1 & 1.8$_{-0.8}^{+0.4}$ & 16$_{-16}^{+10}$ &208$_{-48}^{+81}$  &1.019/1103\\ \nl
NL+Fabian  & 3.18$\pm$0.14 & 1.81$\pm$0.04 & 2.6$\pm$0.2 & 6.37$_{-0.05}^{+0.05}$ & 66$_{-32}^
{+16}$ &  6.4({\it fixed}) & 2.6$_{-0.9}^{+0.7}$ & 39$_{-7}^{+4}$ & 156$_{-47}^{+49}$ & 1.017/1101 \\ \nl
NL+Laor & 3.16$\pm$0.14 & 1.81$\pm$0.04 & 2.6$\pm$0.2 & 6.38$_{-0.04}^{+0.04}$ & 84$_{-30}^{+30}$  &
  6.4({\it fixed}) & 2.4$_{-0.7}^{+0.7}$& 40$_{-9}^{+7}$ & 163$_{-66}^{+84}$  & 1.017/1101 \\ \nl
\enddata 
\tablenotetext{}{fits: \parbox[t]{15cm}{ Fabian -- Disk line model from Fabian et al. (1989);
                       Laor -- Disk line model from Laor (1991); 
                       NL+Fabian -- narrow gaussian plus Fabian model; 
                       NL+Laor --   narrow gaussian plus Laor model.}
                       }
\tablenotetext{a}{SIS normalization at 1 keV in unit of 0.01 photon~cm$^{-2}$~s$^{-1}$~keV$^{-1}$}
\end{deluxetable}

\begin{deluxetable}{cllllllll}
\tablewidth{0pt}
\tablecaption{Results of the soft excess modeling}
\tablehead{
\colhead{fits} & \colhead{$N_H^{(1)}$} & \colhead{$\Gamma$} & \colhead{$N_H^{(2)}$} &
\colhead{kT} & \colhead{abudance} & \colhead{$A_{RS}^{(a)}$} 
& \colhead{$A_{pl}^{(b)}$} & \colhead{$\chi^2_\nu/\nu$} \\
\colhead{} & \colhead{10$^{22}$~cm$^{-2}$} & \colhead{} & \colhead{10$^{22}$~cm$^{-2}$} &
\colhead{keV} & \colhead{Solar unit} & \colhead{ $10^{-4}$ } 
& \colhead{ 10$^{-4}$} & 
}

\startdata
RS & 3.15$\pm$0.08 & 1.82$\pm$0.03 & 0.22$_{-0.13}^{+0.62}$ & 
0.77$_{-0.15}^{+0.12}$ & 0.031$_{-0.014}^{+0.038}$ & 26$_{-10}^{+19}$ & 
\nodata & 1.029/1297 \\ \nl
PL & 3.25$\pm$0.07 & 1.84$\pm$0.03 & 0.06$_{-0.02}^{+0.07}$ & 
\nodata &  \nodata & \nodata & 
3.4$_{-0.6}^{+0.8}$ & 1.033/1299 \\ \nl
RS+PL & 3.22$\pm$0.10 & 1.82$\pm$0.03 & 0.19$_{-0.10}^{+0.26}$ &
0.67$_{-0.41}^{+0.17}$ & 1.0({\it fixed}) & 1.0$_{-0.5}^{+1.5}$ & 
3.5$_{-1.4}^{+1.4}$ & 1.024/1297 \\ \nl
\enddata
\tablenotetext{a}{Raymond-Smith normalization, 
A$_{RS}=(10^{-14}/(4\pi D^2)\int n_en_h$dV, where D is the distance to the source in cm, n$_e$ and n$_h$ are the electron and H densities in cm$^{-3}$.} 
\tablenotetext{b}{the second power-law normalization, in 0.01 photon~cm$^{-2}	$~s$^{-2}$~keV$^{-1}$.}
\tablenotetext{}{fits:\parbox[t]{15cm}{ RS -- Raymone-Smith model with variable abundances, 
PL -- scattered power-law, 
Pl+RS -- scattered power-law plus  Raymond-Smith model with cosmic abundance.}}
\end{deluxetable}

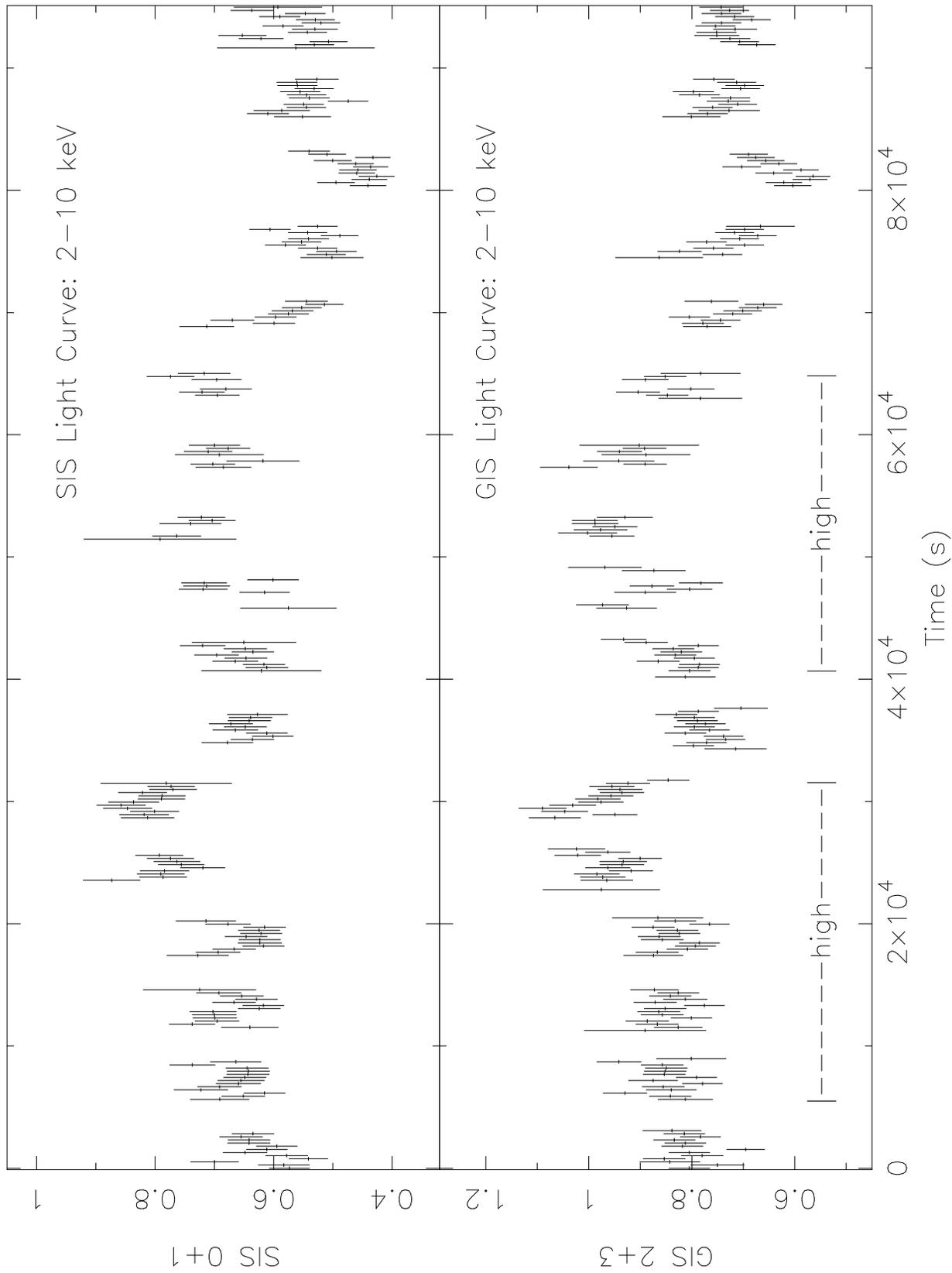
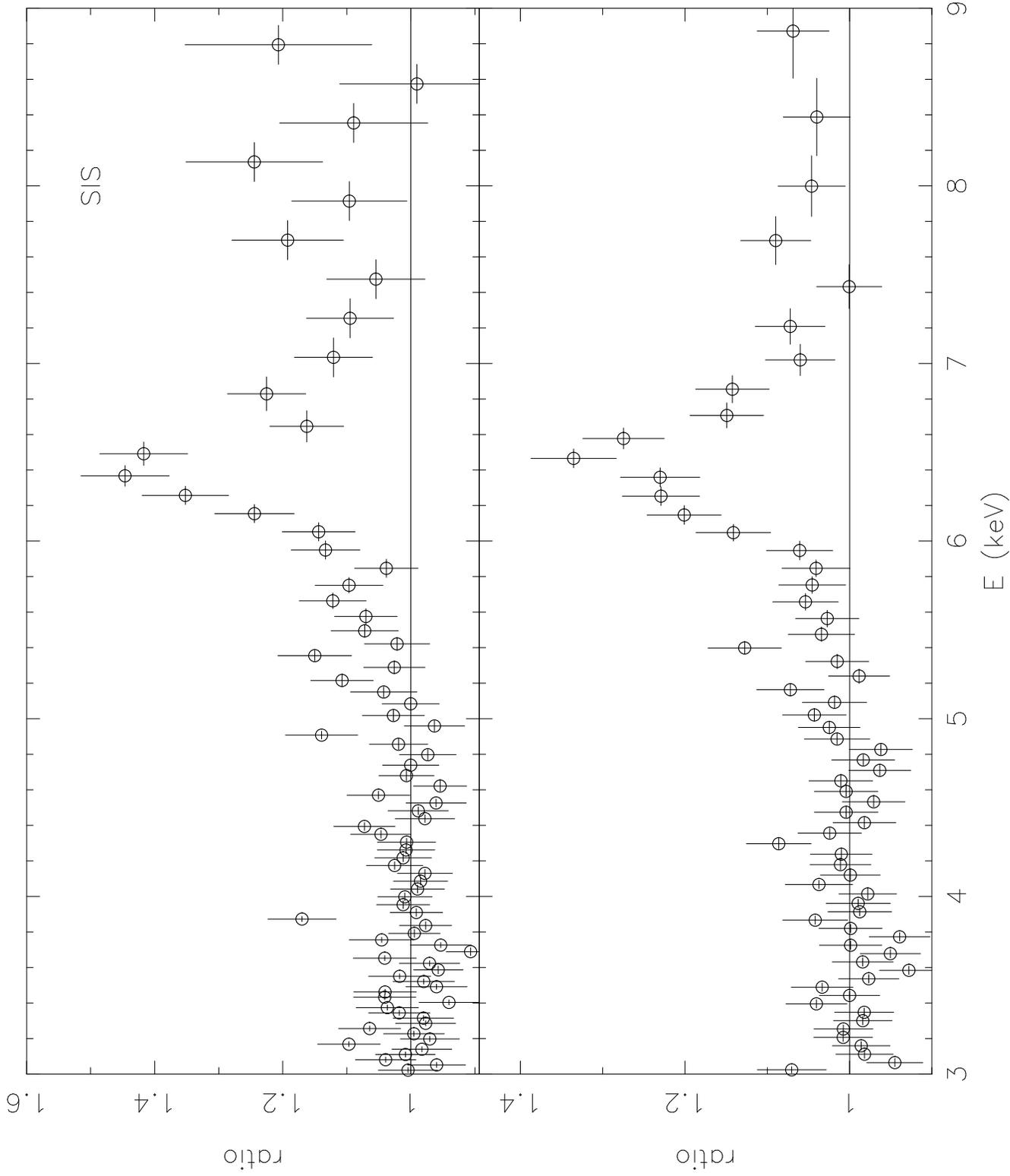
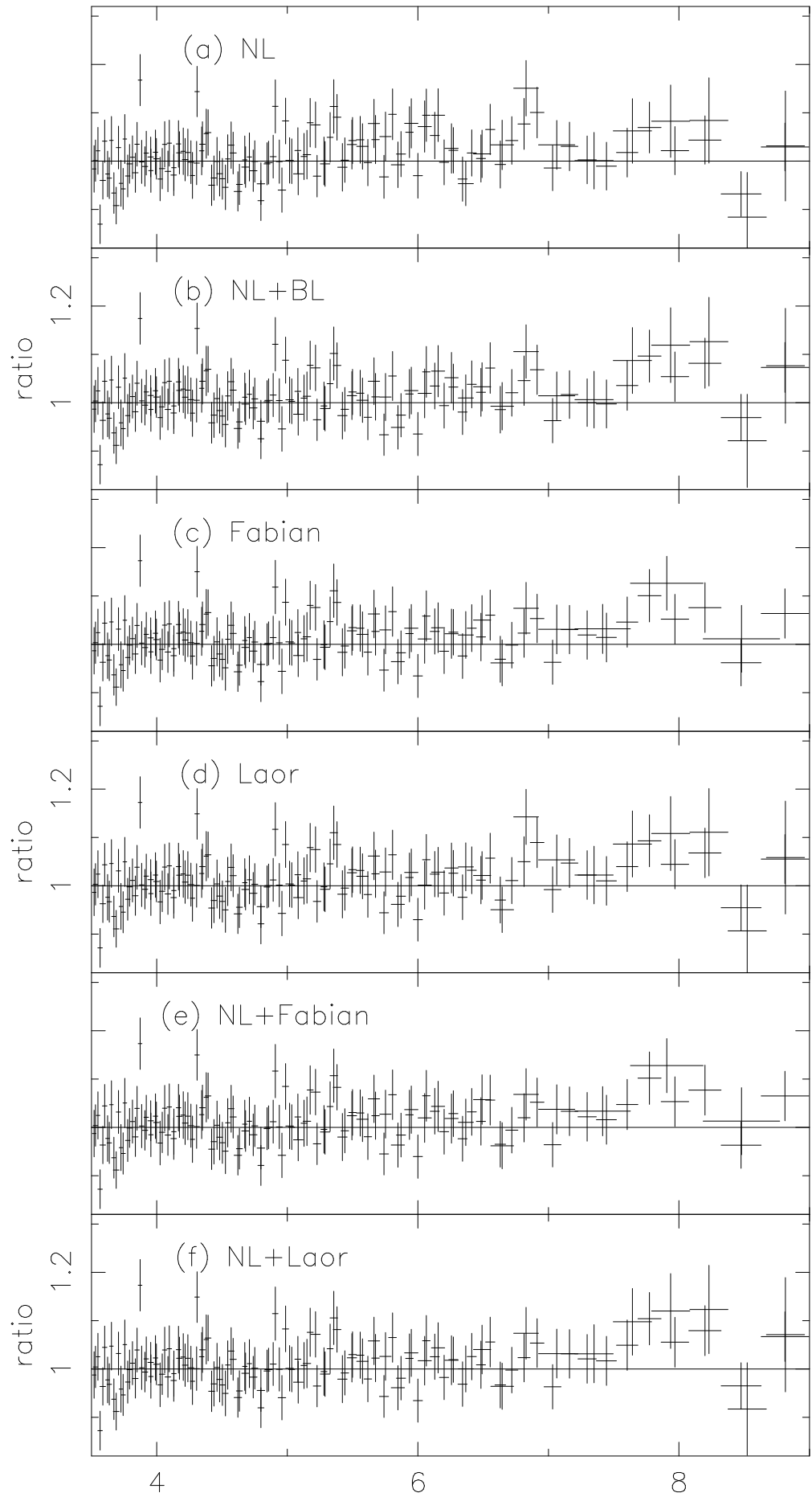
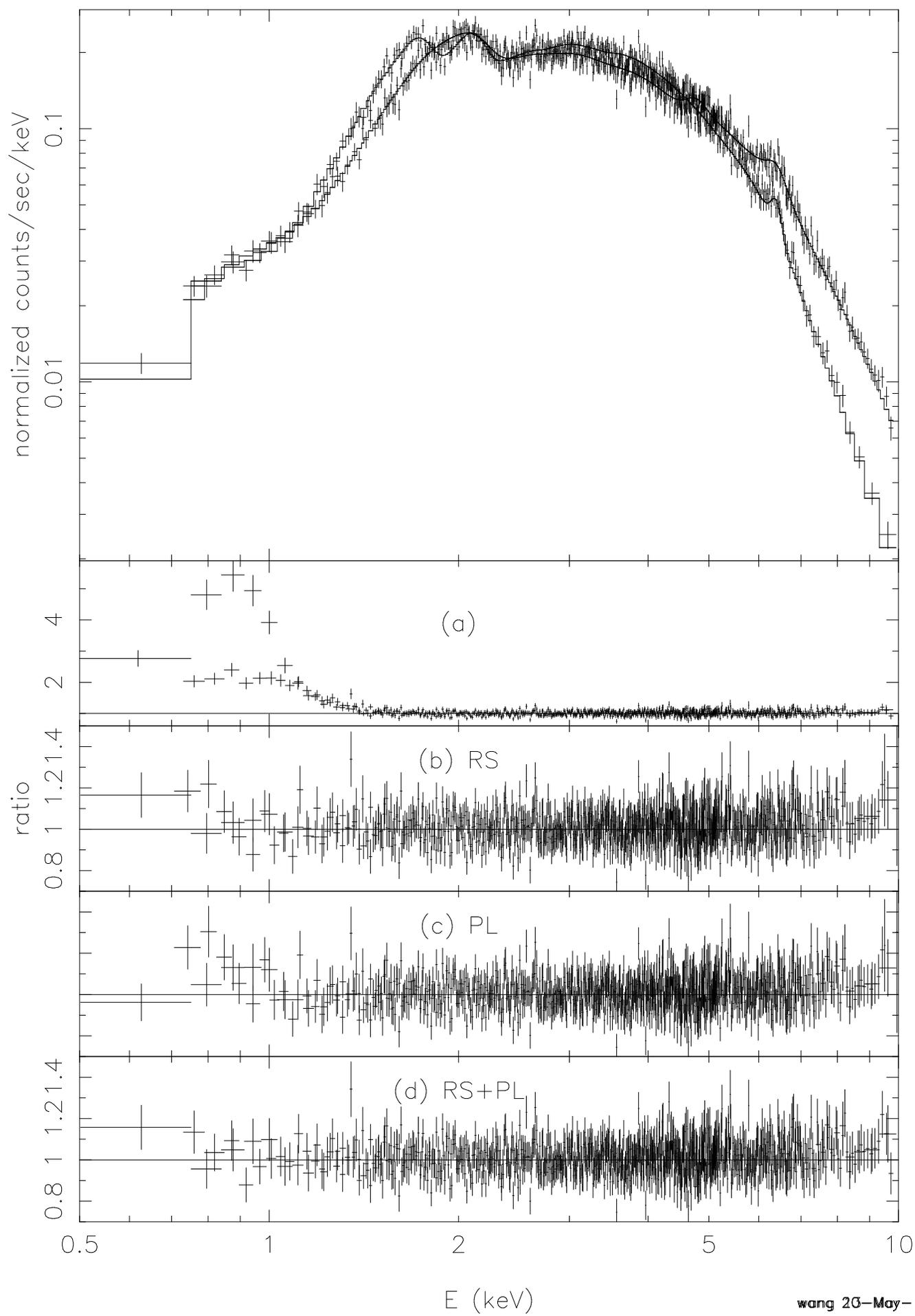
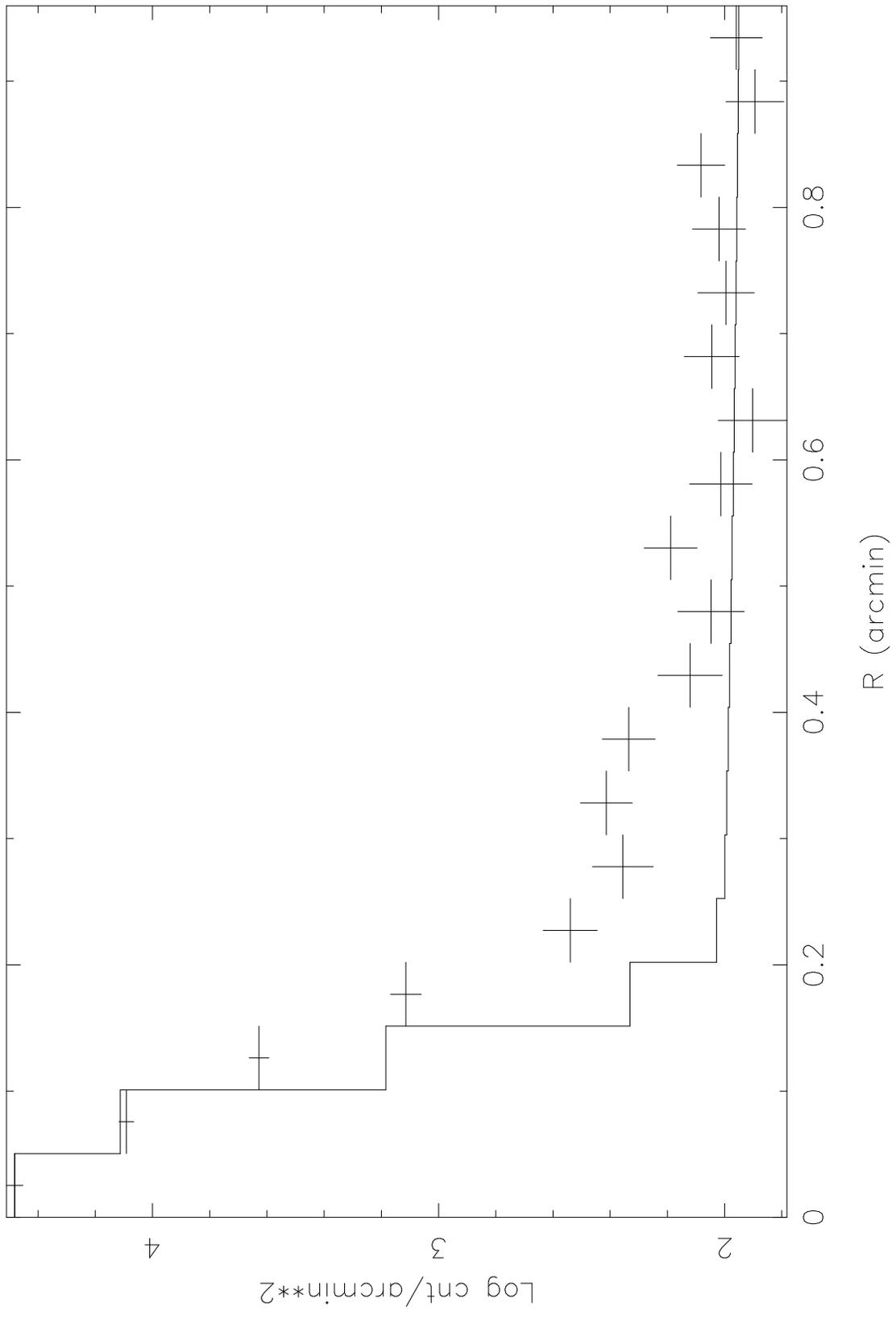
\end{document}